\documentclass[final,3p,times,numbers,sort&compress]{elsarticle}
\usepackage{amsmath,amssymb,amsfonts}
\usepackage{amsthm}
\usepackage{graphicx}
\usepackage{hyperref}

\usepackage{algc}
\usepackage{xcolor}

\definecolor{mycolor}{rgb}{0.7,0.3,0.3}

\usepackage{lineno}

\usepackage{amssymb,amsmath,graphicx}
\usepackage{epigraph}

\def\tr{{\raise0pt\hbox{$\scriptscriptstyle\top$}}}

\begin{document}

\begin{frontmatter}

\title{Symphony of high-dimensional brain. \\ {\small Reply to comments on ``The unreasonable effectiveness of small neural ensembles in high-dimensional brain''}}

\author[LeicMath,NNU]{Alexander N. Gorban\corref{cor1}}
\ead{a.n.gorban@le.ac.uk}
\author[NNU,UCM]{Valeri A. Makarov}
\ead{vmakarov@ucm.es}
\author[LeicMath,NNU,LETI]{Ivan Y. Tyukin}
\ead{i.tyukin@le.ac.uk}

\address[LeicMath]{Department of Mathematics, University of Leicester, Leicester, LE1 7RH, UK}
\address[NNU]{Lobachevsky University, Nizhni Novgorod, Russia}
\address[UCM]{Instituto de Matem\'{a}tica Interdisciplinar, Faculty of Mathematics, Universidad Complutense de Madrid,  28040 Madrid, Spain}
\address[LETI]{Saint-Petersburg State Electrotechnical University, Saint-Petersburg,  Russia}
\cortext[cor1]{Corresponding author}
\end{frontmatter}
\vspace{5mm}

This reply is not a `comment on comments' but rather a `selective echo' that  resonates with some ideas and questions raised by the commentators  \cite{KurkovaComm2019, TozziComm2019, VaronaComm2019, BarrioComm2019, KreimanComm2019, FortunaComm2019, LeeuwenComm2019, QuianQuirogaComm2019, KreinovichComm2019}. We would like to thank all of them. All the individual comments and ideas come together as a symphony of thought, and each individual part changes the way the multidimensional brain is understood. The comments include many deep thoughts  and inspiring questions  and deserve a close reading.

\paragraph{The flight from independence and uniformity} K{\r{u}}rkov{\'a} \cite{KurkovaComm2019} has emphasized that many classical results about measure concentration assume either uniform probability distribution (with respect to the Lebesgue measure in space or the rotationally invariant measures on spheres \cite{Levy1951, Bal1997}),  or independence (`product measures' \cite{Talagrand1995}),  or both. In classical machine learning theory, data are typically assumed to be i.i.d. samples. On the contrary, in real life data samples are neither i.i.d. nor uniformly distributed.

A similar comment about the difference between the theoretical random distributions and ``extremely non-random'' real distributions has been received from Geoffrey Hinton (personal communication to ANG). The obvious gap between the current theory and practice  is a problem, indeed, and the machine learning theory should be revised to deal with non-independent and extremely non-uniformly distributed data. K{\r{u}}rkov{\'a} with Sanguineti proposed their approach to this problem in a very recent publication \cite{KurkovaSang2019}. Our first two stochastic separation theorems used the classical assumptions about  either uniformity of distributions (uniform distribution in balls) or independence (product distributions in cubes)
\cite{GorbTyu2017}. After addressing these basic cases we started our `flight from independence and uniformity' in the paper ``Augmented Artificial Intelligence: a Conceptual Framework'' \cite{GorbanGrechukTykin2018} and developed this work further in \cite{GorbanGolubGrechTyu2018, GorbanMakarovTyukin2019}.

The intermediate conclusion  is as follows. Stochastic separation theorems do not need hypotheses about independence and uniform distribution of data (as well as any other hypothesis about special distributions like the Gaussian one). The main condition we used instead of these simplifications is: sets of small volume should not have a high probability (further specifications in what `small' and `large' mean here can be found, for example in Theorem 1, Definition 3, and Theorem 6 from  \cite{GorbanMakarovTyukin2019}, Theorem 7 and Sec. 6 ``Quasiorthogonal sets and Fisher separability of not i.i.d. data''  in \cite{GorbanGolubGrechTyu2018}). In particular,  instead of uniform or Gaussian distributions general log-concave distributions can be used, and this is just an example. Fisher separability of random points holds for more general distributions that satisfy the SmAC (SMeared Absolute Continuity) condition (see \cite{GorbanMakarovTyukin2019}, Definition 3 or \cite{GorbanGolubGrechTyu2018}, Definition 4). Donoho and Tanner \cite{DonohoTanner2009} observed   `surprising' universality of linear separation in numerical experiments with high-dimensional random datasets. They  proved this property for Gaussian distributions and formulated the problem to determine a general definition of distributions with such separability. We now know that explicit and useful  Fisher’s separability is also typical, and the general definitions of distributions with such properties are understood much better now.

The qualitative essence of these definitions is always the same: {\em small volume sets should not have a high probability}. If such sets exist then an additional challenge arises: extract the sets of small volume and relatively large probability from the data space. The remaining part will satisfy the stochastic separation theorems with Fisher's separability, while the extracted set will have effectively lower dimension and can be approximated by the low-dimensional continua \cite{GorZin2010, GorbanKegl2008, ZinMir2013}.  The extracted low-dimensional subsets can have complex topology and analysis of this topology is important for many modern applications like single-cell omics in bioinformatics \cite{ChenPinelloNatCom2019}.
 Here we reveal a special complementarity principle \cite{GorTyukPhil2018}: the data space is split into a low volume (low dimensional) subset, which requires nonlinear methods for data approximation and analysis, and a high dimensional subset, where the linear methods (Fisher discriminants) work well.

Following the K{\r{u}}rkov{\'a} comment \cite{KurkovaComm2019} and papers  \cite{KurkovaSang2019,  GorbanGrechukTykin2018, GorbanGolubGrechTyu2018, GorbanMakarovTyukin2019}, we invite all readers to join the flight from independence and uniformity. This is a call for revisiting the existing general machine learning theory to bring it closer to real life problems.

\paragraph{Linear algebra vs topology} Tozzi and Peters \cite{TozziComm2019} used the Borsuk--Ulam Theorem (BUT) to explain the blessing of dimensionality. In the topological form (proven by Fet) \cite{Fet1954} BUT is: {\em For every continuous involution $a\mapsto a^*$ ($a^{**}=a$) of a sphere  $\mathbb{S}^n$ and every continuous map $f: \mathbb{S}^n \to \mathbb{R}^n$ there exists a point $x\in \mathbb{S}^n$ such that $f(x)=f(x^*)$.} The most well-known form of BUT uses the central symmetry $a \mapsto -a$ instead of a general continuous involution. The gluing of opposite points is interpreted by Tozzi as a decrease in information in topological sense. Reversely, the dimensionality increase  `unglues' points, and information increases.  In more detail, this point of view is presented and illustrated with applications in the review paper \cite{TozziRev2019}. It is necessary to add that for a continuous map $f: \mathbb{S}^n \to \mathbb{R}^n$ `almost every' point is glued with another point (just imagine a projection of a 2D sphere on a plane). We should stress, however,  that the volume/probability concentration arguments with concentration of the volume near equator are invariant with respect to any orthogonal transformation that preserves the whitened distribution with unit covariance matrix. This property will be destroyed by a typical homeomorphism. On the contrary,  the BUT  in the topological form and other topological statements are invariant with respect to any homeomorphism. Both arguments, topological and linear, are important, but  further work is needed  to clarify the relations between them (at least, for us). Another intriguing problem with interplay of linear algebra and topology is the definition of `intrinsic dimension' of data. Real data never are  i.i.d. sample from a regular and fixed distribution. There are correlations between data points, the distribution (assuming that it exists) can change over time, there may exist a hidden “concept drift” and many other complications. We proposed to use the Fisher separability property for definition of intrinsic dimension of data: it is the dimension of a sphere having the same separability properties for i.i.d. samples from a uniform distribution (the `equivalent sphere') \cite{GorbanGolubGrechTyu2018}. This concept was developed further and tested on biological data \cite{ABZdimension2019}. Zinovyev et al \cite{ABZdimension2019} showed  how the suggested approach can be used to explore the structure of the data
types that are generally considered hard to analyze (mutation and single cell RNA-Seq data).
We expect further interesting and useful discoveries at the boundary between topology and geometry of data.

\paragraph{Communication between neuroscience and Artificial Intelligence (AI)} Varona \cite{VaronaComm2019} discussed relations between machine learning and brain studies, from their common roots to a long period of divergence followed by the modern trend towards convergence. The ``bidirectional communication between machine learning and neuroscience'' is an important way to enrich both disciplines. Problems arising in the ontogeny of the brain and in machine learning can be similar. Both in the brain and in machine learning, re-training large ensembles of neurons is  in fact impossible to realize in many real-life situations/applications. The robust low-dimensional neural dynamics in high dimensional brains is important  for understanding  implementation of cognitive functions  (see, for example, \cite{RabiVaron2018}). However,  there is a difference in this similarity. We agree with Varona that AI is far from incorporating all highly heterogeneous, dynamically robust, and error-handling elements of real brain. Further bi-directional exchange of ideas and methods between AI and neuroscience is needed.

\paragraph{Moving from ``brainland'' to ``flatland'' and backwards} Barrio \cite{BarrioComm2019} has directed our attention to the intriguing interplay between the low and high-dimensional worlds, between the high-dimensional brainland (the whole world of brain) and low-dimensional flatland of small neural ensembles. A flexible walk along the stairs of models of different dimensions is necessary to understand the brain dynamics. In data analysis, we can find a partial realization of this idea in our complementarity principle \cite{GorTyukPhil2018}: extraction of a lower dimensional subset for non-linear analysis \cite{GorZin2010} and application of simple linear methods to essentially high-dimensional part of data \cite{GorbanGolubGrechTyu2018}. Some open access software for this problems is available on github \cite{githubLaboratory}.

\paragraph{Jewels of low-dimensional representation in complexity of high dimensional systems} Kreiman \cite{KreimanComm2019} analyzed three forms of the curse of dimensionality and high-dimensional pitfalls in neuroscience: (1) Huge diversity of possible stimuli and tasks; (2) Huge number of neurons; (3) No adequate tools -- most modeling and analysis tools are developed for small dimensional worlds. Nevertheless, there are many examples in neuroscience, where small  dimensional models  explain a large fraction of a cognitive behavior. Most of them are manifestations of the concentration of measure phenomena. Kreiman attracted attention to the mechanism of removing the curse of dimensionality by projection on random bases. This approach is based on the Johnson and Lindenstrauss lemma and related topics \cite{JohnsonLindenstrauss1984, IndykMotwani1998, GorbTyuProSof2016}. According to this lemma and its probabilistic proof, a set of points in a high-dimensional space can be embedded into a space of much lower dimension in such a way that ratios of distances between the points are nearly preserved. The map used for the embedding  can even be taken to be a random orthogonal projection. This method of random projections was used in a recent publication \cite{GaoEtAlbioRxiv2017} to explain a striking simplicity underlying multi-neuronal data. It is worth to mention that the Johnson and Lindenstrauss lemma and the method of random projections are based on the measure concentration phenomena \cite{GorbTyuProSof2016}. The `brainland' can turn out to be a small dimensional world after all.

\paragraph{Coupling of stochastic separation theory with nonlinear dynamics techniques} Fortuna \cite{FortunaComm2019} proposed to combine the data analysis methods based on statistical physics of data and  advanced experimental based nonlinear dynamics techniques. He demonstrated the power of dynamical models in neuroscience by three examples: (i) a multi jump resonance behavior of  an ensemble of neurons that explains selectivity, (ii) stochastic resonance  in a group of neurons (when the signal to noise ratio at the output of the neuron
ensemble is greater than the input signal to noise ratio), and (iii) synchronization effect in networks in the brain. The challenge is in effective combination of the stochastic separation technique  and topological data approximation with low-dimensional dynamical models and model reduction methods.

\paragraph{Models differ from reality}  van Leeuwen \cite{LeeuwenComm2019} presented an anti-cybernetic manifesto. He sought to show that the classical cybernetic understanding of the brain had  run its course. According to his comment, cybernetic approach to brain study is a misleading metaphor. For example, it does not take into account non-local field interaction of neurons. The brain  is actively generating predictions about what will happen and engages in controlled hallucination. The brain, unlike legacy software, is constantly updating itself. These and other important properties of the brain may deplete the cybernetic approach and make it useless. Such a serious gap between modern AI and real brain can ruin our hope for the efficient communication between neuroscience and AI. According to van Leeuwen, we have to modify our mathematical and computer science approaches so that this communication becomes useful for both parts. What can we say here? First of all, models always differ from reality. The question is not, however, whether there exist important properties of the brain that are neglected by a model. They always exist. The question is whether the model reflects any important aspects of reality. A model is an intelligent device that should perform useful work. Simple models, ranged from stochastic separation, concept cells, and Hebb's rule described in our review \cite{GorbanMakarovTyukin2019} to direct simulation of large networks of relatively simple neurons \cite{Izhikevich2008} and rule based models of development neuronal network generating  behavior \cite{RobertsBorisyuk2014}, have not yet exhausted their capabilities.  Nevertheless, we are ready to accept the expert opinion of van Leeuwen and agree that a new conceptual basis of brain modelling is needed.  Let us indeed hope that a new theoretical framework will be eventually developed in neuroscience that will appeal to mathematician and computer scientists, and
 there will be a decent place for simple basic models.

\paragraph{Akaki Akakievich and the modern point of view on Concept Cells} Quian Quiroga \cite{QuianQuirogaComm2019} clarified and commented on modern ideas about Concept Cells and sparse coding in the form of a revised Lettvin story about grandmothers cells. He stressed that
Concept Cells are not involved in identifying a particular stimulus or concept. They are
rather involved in creating and retrieving associations and can be seen as the ``building
blocks of episodic memory''. We analyzed basic dynamical models of such associations \cite{TyukinBrain2017} (see also \cite{GorbanMakarovTyukin2019}, Sec. 4. ``Encoding and rapid learning of memories by single neurons'') but this analysis should be extended and compared quantitatively to experimental data.  The professional and clearly presented explanation of the basic concepts in Quian Quiroga's commentary is a very useful addition to our review, and we are happy to recommend this commentary to all readers.

\paragraph{``Happiness is when you are understood''} The beautiful text of Kreinovich \cite{KreinovichComm2019} does not need comments.  It is our pleasure just to mention that in our internal discussions of the manuscript \cite{GorbanMakarovTyukin2019}, we have many times quoted Pasternak's poetry about the ``heresy of unheard-of simplicity''. Kreinovich found the same golden words of Pasternak to describe the apparent simplicity in a multidimensional brain. This is an example of an amazing intellectual resonance.

\section*{Acknowledgments}
The work was supported by the Ministry of Science and Higher Education of the Russian Federation (Project No. 14.Y26.31.0022). Work of ANG and IVT was also supported by Innovate UK (Knowledge Transfer Partnership grants KTP009890; KTP010522) and University of Leicester. VAM acknowledges support from the Spanish Ministry of Economy, Industry, and Competitiveness (grant FIS2017-82900-P).

 \end{document}